\documentclass[aps,twocolumn,groupedaddress]{revtex4}

\usepackage{graphicx} 

\begin{document}
\title{ LIMITING SENSITIVITY OF SUPERCONDUCTIVE LC - CIRCUIT
\\ PLACED IN WEAK GRAVITATIONAL WAVE}

\author{ P.~Fortini$^{a}$, G.~N.~Izmailov$^{b}$ \\}

\affiliation{$^{a}$Dipartimanto di Fisica, Universit\`a di Ferrara and 
INFN, Sezione di Ferrara, 40100 Ferrara, Italy \\
$^{b}$ Department of Applied Mathematics and Physics, Moscow State Aviation 
Institute (Technical University), Volokolamskoe sh., 4 Moscow 125871 Russia}
\date{\today}

\begin{abstract}
It is demonstrated that the sensitivity of a superconductive LC - circuit 
placed in a weak gravitational wave is limited by two factors. One is the 
quantization of the magnetic flux through the circuit, the second one is the
fraction of the elementary charge (effect Laughlin - Stormer - Tsui).
Application to a possibility of using a superconductive LC - circuit as a weak 
gravitational waves detector is discussed.

\vspace{0.5cm}

PACS number(s): 04.80
\end{abstract}


\maketitle

\vspace{0.5cm}
 
The attractive goal of a gravitational astronomy stimulates us to enlarge
the set of gravitational wave antennae based on different physical
principles. Because it is evident that the diversity of experimental
devices can help us to get complete and reliable information.  

 As it was proved in  \cite {Fo}, if a RLC - circuit is placed in a wave
front plane of a monochromatic gravitational wave (GW) electromagnetic 
oscillations can be, so the RLC - circuit is a gravitational wave 
detector. The relative position of RLC - circuit elements and the 
gravitational wave front on fig.1 is shown.

\begin{figure*}[t]
\begin{center}
\includegraphics[scale=0.12]{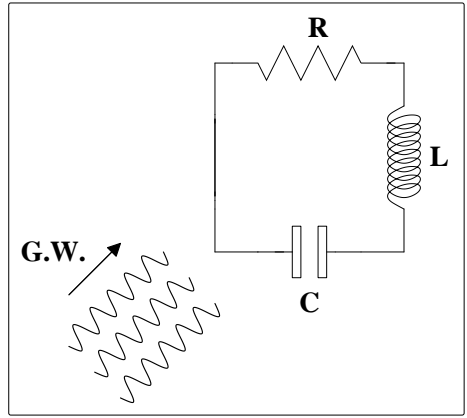}
\caption{A relative disposition between the plane of GW and the plane 
of RLC-circuit }
\label{uno}
\end{center}
\end{figure*}
 
 The incoming GW changes mutual positions of circuit conductors and its
 cross - sections, consequently, magnitudes of resistor, inductance and
capacity are changes also in a phase with GW. Its variations are
proportional to the GW amplitude. In \cite{Fo} obtained that in orthogonal 
TT frame (in that frame the RLC - circuit is stated relatively to the GW front)
 \begin{equation}
  \left\{
   \begin {array}{rcl}
    R(t)&=&R~[1~+~\varepsilon_{R}(t)] \ ,\\ [5pt]
    L(t)&=&L ~[1~+~\varepsilon_{L}(t)] \ ,\\[5pt]
    \frac {\displaystyle 1}{\displaystyle C_{i}(t)}&=&
    \frac {\displaystyle 1}{\displaystyle C_{i}}~
    [1~-~\varepsilon_{C_{i}}(t)] \ , \ \ \ \ \ i~=~1,~2 \ ,
   \end {array}
  \right.
 \end{equation}
 where  $R,~L,~C_{i}$  are initial magnitudes
 \begin{equation}
 \varepsilon_{\flat}(t)~=~\epsilon_{{\flat}_{+}}h_{+}(t)~
 +~\epsilon_{{\flat}_{\times}}h_{\times}(t)\ ,
 \end{equation}
 and $g_{\mu,\nu}~=~\eta_{\mu,\nu} + h_{\mu,\nu}$, $\epsilon$ is a
 parameter, $\flat =~R,~L,~C_{i}$, \ $+$ or $\times$  signs of the
 monochromatic GW polarization. For the monochromatic GW we suppose
 \ $ h_{+, \times}(t)~=~ A_{+, \times}\cos(\omega_{g}t + \phi_{+, \times})$.
 
 The assumptions for deriving Eqs. (1) are as follows. The oscillations
 frequency is less then $c/d$ ($c$ is the velocity of light, $d$ is the maximum
 length of the circuit), the field of the GW is uniform or $\lambda_{g}\gg d$
 ($\lambda_{g}$ is the length of the GW).
 
 In these assumptions one can write the equations for oscillations of a
 charge $Q$ in the RLC - circuit in the GW presence
 \begin{equation}
 \ddot Q~+~2\gamma\dot Q~+~\omega_{0}^{2}~Q~=~
 v_{+} h_{+}(t)~+~v_{\times} h_{\times}(t) \ ,
 \end{equation}
 where
 $$
  \begin {array}{rcl}
    v_{+~,\times}&=&\omega_{0}^{2}Q_{0}\displaystyle
        \frac {a}{\displaystyle (a~+~1)^{2}}
                [\epsilon_{C_{1_{+, \times}}}~-~\epsilon_{C_{2_{+,
                  \times}}}] \ ,                                 \\ [6pt]
  \omega_{0}^{2}&=&\displaystyle \frac{a~+~1}{a}\frac 1{LC } \ , \\ [6pt]
  Q_{1}&=&aQ_{2}~=~\displaystyle \frac{a}{a~+~1}~ Q_{0} \ ,      \\ [6pt]
  C_{1}&=&aC_{2}~=~aC \ .                            \\
  \end {array}
 $$
 Eq. (3) in $|\varepsilon_{\flat}|\ll 1$ and $|\dot\varepsilon_{\flat}|\ll
 1$ approximations is derived.
 
 So, the most prominent effect of the action of the G~W is the existence
 of a electromotive force in the RLC - circuit. Its value is defined by
 an amplitude of the GW. In order to simplify following calculations
 it is supposed that only "plus" polarization of the GW is acting on the
 circuit. Plates of the condensers  $C_{1}$ and $C_{2}$ are mutially
orthogonal,
 and $C_{1}~=~C_{1}~=~C$. So $a~=~1$, $v_{+}~=~1/2\omega_{g}^{2}Q_{0}$.
 
 It is obvious in a case of resonance $\omega_{0} = \omega_{g}$ that losses
 in the RLC - circuit confine an amplitude of electric oscillations. In usual
 RLC - circuit with the temperature $T$ its sensitivity is restricted by a 
termal noise
 \begin{equation}
    A_{n}~=~\frac {2}{U_{0}}\sqrt{\frac{\kappa T}{{\cal Q}C\omega_0\tau}}
\ ,
 \end{equation}
 if one suppose $U_{0} = Q_{0}/C,~\kappa$  is the Boltzman con\-stant,
 $\cal Q$ is the quality factor $\cal Q$=$\omega_0/4\gamma$, $\tau$  is
the interval of registration. For $U_{0}~=~10^{5}~V$,
$T~=~4~K$,$ ~\omega_0~=~120\pi$, $ {\cal Q}~=~10^{3}~$, $\tau~=~10^{7}~s$, then 
the minimal amplitude of GW is \cite{Fo}
 \begin{equation}
    A_{n}~=~ 7,6\cdot10^{-22}\ .
 \end{equation}
 
 Let us discuss the sensitivity of the superconductive LC - circuit. We can
define the sensitivity of such circuit from the following reasons.
The magnetic flux is quantized in the
superconductive inductance $L$, and the value of minimum flux is \cite{2}
 \begin{equation}
   \Phi_{0}~=~\frac {\pi \hbar}{e}~=~2.07\cdot10^{-15} ~Wb \ .
 \end{equation}
 According to Faraday's law, if the magnetic flux is changed on
 $\Delta\Phi~=~
 \Phi_{0}$ in the time interval $\Delta t~=~T_{g}/2~(T_{g}~=~2\pi/\omega_{g})$,
 then the electromotive force is created
 \begin{equation}
   E_{L}~=~-\frac{\omega_{g}}{\pi}\Phi_{0}\ .
 \end{equation}
 This electromotive force in the LC~-~circuit is compensated by changes of
 $C_{1}$ and $C_{2}$ voltages (because Kirchgoff's law). The last are
 created by the GW action. As one can conclude
 \begin{equation}
   \Delta U~=~U_{0} A {\cal Q}_s \ ,
 \end{equation}
 where ${\cal Q}_s$ is the quality factor of the LC~-~circuit 
( see [1] p. 57 ), $A$ is an
 amplitude of the GW. So, using Kirchgoff's law, one can have
 \begin{equation}
   E_{L}~+~\Delta U~=~0\ .
 \end{equation}
 
 After substitution (7) and (8) in (9) we find
 \begin{equation}
   \frac{\omega_{g}\Phi_{0}}{\pi}~=~U_{0} A {\cal Q}_s \ .
 \end{equation}
 
 Consequently, the boundary value of the GW amplitude, which can be detected in the
 LC~-~circuit is
 \begin{equation}
   A_{min_{s}}~=~\frac{\omega_{g}\Phi_{0}}{\pi U_{0} {\cal Q}_s}\ ,
 \end{equation}
 or in another terms
 \begin{equation}
   A_{min_{s}}~=~\frac{\hbar \omega_{g}}{{\cal Q}_s e U_{0} }\ .
 \end{equation}
 
 If one gets the above mentioned values and supposes that ${\cal Q}_s~=~10^{6}$
 and $\omega_{g}~=~120\pi$, then
 \begin{equation}
    A_{min_{s}}~=~2,4\cdot10^{-24}\ .
 \end{equation}
 
As it was shown recently \cite{3}, if one lowers the temperature in 
the $LC$-circuit less than $0.4 ~K$ and gets the external magnetic field 
more than $10~T$,
he can use the fraction of the elementary charge $e$ in the experimental 
measurement.  Consequently, we have to rewrite the
formula (12) so that this effect can be included.
It is easy to show that with such technology we can get the sensitivity
 \begin{equation}
   A_{min_{s}}~=~\frac{\hbar \omega_{g}}{{\cal Q}_s e U_{0} }n\ ,
\end{equation}
where $n$ is a fractional part of the elementary charge ( in particular 
1/3, 1/5 and so on).

In conclusion, we notice that the sensitivity of a such device, being 
comparable with a laser interferometer one, can be exploited as an 
effective gravitational wave detector. 

One of the authors (G.~N.~I.) should like to thank the 
INFN - Sezione di Ferrara for financial support of his staying in Ferrara.

\end{document}